\documentclass[10pt,conference]{IEEEtran}

\usepackage{amssymb}
\usepackage{amsmath}
\usepackage{eucal}	

\newtheorem{theorem}{Theorem}[section]

\newcommand{\defeq}{\stackrel{\triangle}{=}}

\begin{document}

\title{Parametrized Stochastic Grammars for\\
RNA Secondary Structure Prediction}
\author{\authorblockN{Robert S. Maier}
\authorblockA{Departments of Mathematics and Physics\\
University of Arizona\\
Tucson, AZ 85721, USA\\
Email: rsm@math.arizona.edu}}

\maketitle

\begin{abstract}
We propose a two-level stochastic context-free grammar (SCFG) architecture
for parametrized stochastic modeling of a family of RNA sequences,
including their secondary structure.  A~stochastic model of this type can
be used for maximum a~posteriori estimation of the secondary structure of
any new sequence in the family.  The proposed SCFG architecture models RNA
subsequences comprising paired bases as stochastically weighted
Dyck-language words, i.e., as weighted balanced-parenthesis expressions.
The length of each run of unpaired bases, forming a loop or a bulge, is
taken to have a phase-type distribution: that of the hitting time in a
finite-state Markov chain.  Without loss of generality, each such Markov
chain can be taken to have a bounded complexity.  The scheme yields an
overall family SCFG with a manageable number of parameters.
\end{abstract}

\section{Introduction}
\label{sec:intro}
In biological sequence analysis, probability distributions over finite
($1$-dimensional) sequences of symbols, representing nucleotides or amino
acids, play a major role.  They specify the probability of a sequence
belonging to a specified family, and are usually generated by Markov
chains.  These include the stochastic finite-state Moore machines called
hidden Markov models (HMMs); or infinite-state Markov chains such as
stochastic push-down automata (SPDAs).  By computing the most probable path
through the Markov chain, one can answer such questions as ``What hidden
(e.g., phylogenetic) structure does a sequence have?'', and ``What
secondary structure will a sequence give rise~to?''.  The number of Markov
model parameters should ideally be kept to a minimum, to facilitate
parameter estimation and model validation.

The a~priori modeling of an RNA sequence family is considered here.  Due~to
Watson--Crick base pairing, a recursively structured RNA sequence will
fold, and display secondary structure.  To model stochastically both
pairings and runs of unpaired bases (which form loops and bulges), results
from a subfield of formal language theory, the {\em structure theory of
weighted strings\/}~\cite{KuichSalomaa} (each string being weighted by an
element of a specified `semiring' such as~${\mathbb{R}}_+$), are reviewed
and employed in stochastic model construction.

In Section~\ref{sec:duration}, {\em duration modeling\/} is discussed: the
modeling of a probability distribution on `runs', i.e., on the natural
numbers~$\mathbb{N}$.  A non-RNA biological example is the modeling and
prediction of CpG~islands in a DNA sequence.  A~sequence may flip between
CpG and non-CpG states, with distinct HMMs for generation of symbols in
$\{A,T,G,C\}$.  For ease of HMM parameter estimation, and for finding the
most probable parse, or path through the model (e.g., by the Viterbi
algorithm), the length of each CpG island and non-CpG region should be
modeled in a Markovian way, as the first hitting time in a finite-state
Markov chain.  That~is, on~$\mathbb N=\{0,1,2,\dots\}$, the set of possible
lengths, it should have a {\em phase-type
distribution\/}~\cite{Neuts,OCinn90}.  There is a theorem of the author's
on such distributions~\cite{Maier8a}, which grew out of results on
positively weighted {\em regular\/} sequences~\cite{Katayama,Soittola}.  It
says that without loss of generality, the structure of the Markov chain can
be greatly restricted: its `cyclicity' can be required to be at most~$2$.
This has implications for HMM parametrization.

The generating function $G(z)$ of a phase-type (PH) distribution
on~$\mathbb N$ (which is a normalized $\mathbb{R}_+$-weighted regular
language over a $1$-letter alphabet) is a {\em rational\/} function of~$z$.
Going beyond regular languages to the context-free case yields an {\em
algebraic\/} generating function: one of several variables, if each type of
letter in the sequence is separately kept track~of.  In RNA secondary
structure prediction, stochastic context-free grammars (SCFGs), usually in
Chomsky normal form, have been used~\cite{Sakakibara94}.  They tend to be
complicated; if the grammar has $k$~non-terminal symbols, then it may have
$O(k^3)$~transition probabilities, which must be estimated from training
sequences~\cite{Lari90}.  What is needed is a class of SCFGs with
(i)~restricted internal structure, (ii)~equivalent modeling power, and
(iii)~computationally convenient parametrization.  Finding such a class of
models is a hard problem: even on the level of $1$-letter-alphabet (i.e.,
univariate) generating functions, it involves the constructive theory of
algebraic functions.

In Section~\ref{sec:algebraic}, as a small step toward solving this
problem, it is pointed~out that there is a class of probability
distributions on~$\mathbb N$ with generating functions (i.e.,
$z$-transforms) that are algebraic and non-rational, which can be
conveniently parametrized.  This is the class of algebraic {\em
hypergeometric distributions\/}.  E.g., the $\mathbb{N}$-valued random
variable~$\tau$ could satisfy $\sum_{n=0}^\infty z^n\, Pr(\tau=n)\propto
{}_2F_1(a,b;c;z)$, where ${}_2F_1(a,b;c;\cdot)$ is Gauss's (parametrized)
hypergeometric function.  If $a,b,c$ are suitably chosen, $n\mapsto
Pr(\tau=n)$ will be a probability density function with an algebraic
$z$-transform.  Algebraic hypergeometric probability densities satisfy nice
recurrence relations, and SCFG interpretations for them can be worked~out.

A more general approach toward solving the above problem, not restricted to
the case of a $1$-letter alphabet, employs SCFGs with a two-level
structure.  In~effect, these are SCFGs wrapped around HMMs.  The following
is an illustration.  A~probabilistically weighted Dyck language over the
alphabet $\{a,b\}$, i.e., a distribution over the words in $\{a,b\}^*$ that
comprise nested $a$\textendash\nobreak$b$ pairs, is generated from a
symbol~$S$ by repeated application of the production rule $S\mapsto
p_1\cdot ab+p_2\cdot abS+p_3\cdot aSb +p_4\cdot aSbS$.  The
probabilities~$p_i$ sum to~$1$.  If each of $a,b$ in~turn represents a
weighted {\em regular\/} language over some alphabet~$\Sigma$ (e.g.,~a
PH-distribution if $\Sigma$~has only one letter), then the resulting
distribution over words in~$\Sigma^*$ comes from a SCFG with the stated
two-level structure.  This setup is familiar from (unweighted) language
theory applied to compilation: the top-level structure of a program is
specified as a word in a context-free language, and islands of low-level
structure (e.g., identifier names and arithmetic literals) as words in
regular languages.

In Section~\ref{sec:modeling}, it is indicated how the idea of a SCFG
wrapped around HMMs can be applied to RNA structure prediction: initially,
to the parametric stochastic modeling, in a given sequence family, of the
recursive primary structure that induces secondary folding.  The goal is
parameter estimation and model validation, by comparison with data on real
RNA sequences.  Knudsen and Hein~\cite{Knudsen99} and
Nebel~\cite{Nebel2004} have worked on this, using Dyck-like languages, but
stochastic modeling using distinct SCFG and HMM levels is a significant
advance.

On the level of primary RNA structure, paired nucleotides will make~up a
subsequence of the full nucleotide sequence, and must constitute a Dyck
word, for simplicity written as a word over~$\{a,b\}$.  A distribution over
the infinite family of such Dyck words is determined by the above
stochastic production rule, the parameters $p_1,p_2,p_3,p_4$ in which are
specific to the family being modeled.  The production rule for {\em full\/}
sequences, including unpaired nucleotides, will have not $ab$, $abS$,
$aSb$, $aSbS$ on its right-hand side, but rather $IaIbI$, $IaIbS$, $IaSbI$,
$IaSbS$, where each~$I$ expands to a `run' of unpaired nucleotides.  If the
four nucleotides are treated as equally likely in this context, each~$I$
will be a stochastic language over a $1$-letter alphabet, and the length of
each run is reasonably modeled as having a PH-distribution.  The PH class
includes geometric distributions, but is more general.  The overall SCFG is
obtained by wrapping the Dyck SCFG around the finite-state Markov chains
that yield the PH-distributions.

From a given family of RNA sequences, Dyck SCFG parameters can be
estimated, e.g., by the standard Inside--Outside Algorithm~\cite{Lari90};
and then HMM parameters (i.e., PH-distribution parameters) can be estimated
separately.  By employing a large enough class of PH-distributions, it
should be possible to produce a better fit to data on secondary structure
than were obtained from the few-parameter models of Knudsen and
Hein~\cite{Knudsen99} and Nebel~\cite{Nebel2004}.  Once the family has been
modeled, the most likely parse tree for any new RNA sequence in the family
can be computed by maximum a~posteriori estimation, using the CYK
algorithm~\cite{Sakakibara94}.  The sequence is predicted to have the
secondary structure represented by that parse tree.

\section{Duration Modeling}
\label{sec:duration}
Since loops and bulges in RNA secondary structure comprise runs of unpaired
nucleotides, they can be modeled without taking long-range covariation into
account.  The appropriate stochastic model is an HMM {\em with
absorption\/}, since the accurate modeling of run lengths is a goal.  Any
such HMM will specify a probability distribution on the set of finite
strings~$\Sigma^*$, where $\Sigma=\{A,U,G,C\}$ is the alphabet set, and
long words are exponentially unlikely.  There should be little change in
the nucleotide distribution along typical runs, so the distribution of the
string length $\tau\in\mathbb{N}$ is what is important.

The time~$\tau$ to reach a final (absorbing) state in an irreducible
discrete-time Markov chain on a state space $Q=\{1,\dots,m\}$, with a
transition matrix $\mathbf{T}=(T_{ij})_{i,j=1}^m$ that is {\em
substochastic\/} (i.e., $\sum_{j=1}^m T_{ij}\le1$), and an initial state
vector $\mathbf{\alpha}=(\alpha_i)_{i=1}^m$ that is also substochastic
(i.e., $\sum_{i=1}^m\alpha_i\le1$), is said to have a discrete
PH-distribution.  The substochasticity of $\mathbf{T}$
and~$\mathbf{\alpha}$ expresses the absorption of probability, since they
can be extended to a larger state space $\tilde Q=Q\cup\{m+1\}$, on which
they will be stochastic.  The added state $m+1$ is absorbing.

There is a close connection between PH-distributions and finite automata
theory, in particular the theory of rational series over
semirings~\cite{KuichSalomaa}.  If $A$~is a semiring (a~set having binary
addition and multiplication operations, $\oplus$~and~$\odot$, each with an
associated identity element; but not necessarily having a unary negation
operation), then an $A$-{\em rational sequence\/}, $a=(a_n)_{n=0}^\infty\in
A^{\mathbb{N}}$, is a sequence of the form $\oplus_{i,j=0}^m
\left[u_i\odot(\mathbf{M}^n)_{ij}\odot v_j\right]$, where for some $m>0$,
$\mathbf{M}\in A^{m\times m}$ and $\mathbf{u},\mathbf{v}\in A^m$.  It is an
$A$-weighted regular language over a $1$-letter alphabet.  Semirings of
interest here include $\mathbb{R}$, $\mathbb{R}_+=\{x\in\mathbb{R}\mid
x\ge0\}$, and the Boolean semiring $\mathbb{B}=\{0,1\}$.

\smallskip
\begin{theorem}[\cite{Maier9}]
\label{thm:normalization}
  Any PH-distribution on~$\mathbb{N}$ is an $\mathbb{R}_+$-rational
  sequence.  Any {\em summable\/} $\mathbb{R}_+$-rational sequence, if
  normalized to have unit sum, becomes a PH-distribution.
\end{theorem}

\smallskip
If $\tau\in\mathbb{N}$ is PH-distributed, it is useful to focus on its
$z$-transform, i.e., $G(z)=E\left[z^\tau\right]=\sum_{n=0}^\infty
z^n\,Pr(\tau=n)$.  This will be a rational function, in~$\mathbb{R}_+(z)$.
If the distribution is {\em finitely supported\/}, it will be a polynomial,
in~$\mathbb{R}_+[z]$.

\smallskip
\begin{theorem}[\cite{Maier9}]
\label{thm:2}
  Any PH-distribution on~$\mathbb{N}$ can be generated from finitely
  supported distributions by repeated applications of (i)~the binary
  operation of mixture, i.e., $G_1,G_2\mapsto pG_1+(1-p)G_2$, where
  $p\in(0,1)$, (ii)~the binary operation of convolution, i.e.,
  $G_1,G_2\mapsto G_1G_2$, and (iii)~the unary `geometric mixture'
  operation, i.e., $G\mapsto (1-p)\sum_{k=0}^\infty p^kG^k=(1-p)/(1-pG)$,
  where $p\in(0,1)$.
\end{theorem}

\smallskip
This is a variant of the Kleene--Sch\"utzenberger Theorem on the
$A$-rational series associated to $A$-finite automata~\cite{KuichSalomaa}.
The Boolean ($A=\mathbb{B}$) case of their theorem is familiar from formal
language theory: it says that any regular language over a finite alphabet
can be generated from {\em finite\/} languages by repeated applications of
(i)~union, (ii)~concatenation, and (iii)~the so-called Kleene star
operation.  Just as in formal language theory, the third operation of
Theorem~\ref{thm:2} can be implemented on the automaton level by adding
`loopback', or cycle-inducing, transitions from final state(s) back to
initial state(s).

\smallskip
\begin{theorem}[\cite{Maier8a}]
\label{thm:3}
The unary--binary computation tree leading to any PH-distribution
on~$\mathbb{N}$, the leaves of which are finitely supported distributions,
can be required without loss of generality to have at most~$2$ unary
`geometric mixture' nodes along the path extending to the root from any
leaf.  That~is, those operations do~not need to be more than doubly nested.
\end{theorem}

\smallskip
This is a normalized, or `stochastic', version of a result on the
representation of $\mathbb{R}_+$-rational
sequences~\cite{Katayama,Soittola}.  Results of this type are strongly
semiring-dependent.  It is not difficult to see that in the cases
$A=\mathbb{R}$ and~$\mathbb{B}$, the analogue of the number~`$2$' is~`$1$'.
(This is because, e.g., a $\mathbb{B}$-rational sequence is simply an
sequence in~$\mathbb{B}^\mathbb{N}$ that is eventually periodic.)  The
proof of Theorem~\ref{thm:3} is an explicit construction, which respects
positivity constraints at each stage.  That~is, the construction solves the
{\em representation problem\/} for univariate PH-distributions, which has
strong connections to the positive realization problem in control
theory~\cite{Commault2003}.

What the theorem says, since operations of types (i),(ii),(iii) correspond
to parallel composition, serial composition, and cyclic iteration of Markov
chains, is that any PH-distribution arises without loss of generality from
a Markov chain in which cycles of states are nested at most $2$~deep.
That~is, the chain may include cycles, and cycles within cycles, but not
cycles within cycles within cycles.  So for modeling purposes, the chain
transition matrix~$\mathbf{T}$ may be taken to have a highly restricted
structure.  A~completely connected transition graph on a state space of
size~$m$ would have ${m}^{2}$~possible transitions, and would be
unnecessarily general when $m$~is large.

Unpaired nucleotide run lengths in RNA are naturally modeled as having
(discrete-time) PH~distributions because of the close connection with HMMs,
and the consequent ease of parameter estimation.  However, the class of
PH~distributions is so versatile that it would be useful in this context,
regardless.  Discrete PH~distributions include geometric and negative
binomial distributions, and are dense (in~a suitable sense) in the class of
distributions $Pr(\tau=n)$, $n\in\mathbb{N}$, which have leading-order
geometric falloff as~$n\to\infty$.

Any PH distribution on~$\mathbb{N}$ has a $z$-transform
$G(z)=E\left[z^\tau\right]$ that is rational in the conventional sense;
equivalently, it must satisfy a finite-depth recurrence relation of the
form $\sum_{k=0}^N c_k\,Pr(\tau=n+k)=0$.  In~fact, any probability
distribution on~$\mathbb{N}$ with (i)~a rational $z$-transform $G(z)$, and
(ii)~the property that the pole which $G(z)$ necessarily has at~$z=1$ is
the {\em only\/} pole on the circle $\left|z\right|=1$, is necessarily a PH
distribution~\cite{OCinn90}.  This is a sort of converse of the
Perron--Frobenius Theorem.  However, there are distributions
on~$\mathbb{N}$ which satisfy~(i) but not~(ii), and are not PH
distributions.  They are necessary $\mathbb{R}$-rational sequences, but are
not $\mathbb{R}_+$-rational sequences as defined above, even though they
are sequences of elements of~$\mathbb{R}_+$ (probabilities).  In
abstract-algebraic terms, this situation is possible because the
semiring~$\mathbb{R}$ is not a {\em Fatou extension\/} of the
semiring~$\mathbb{R}_+$~\cite{KuichSalomaa}.  The existence of pathological
examples of this type does not vitiate the usefulness of PH distributions
in run-length modeling.

\section{Algebraic Sequences}
\label{sec:algebraic}
Most work on RNA secondary structure prediction that draws on formal
language theory has employed SCFGs~\cite{Knudsen99,Nebel2004,Sakakibara94}.
A~CFG in Chomsky normal form~(CNF), used for generating strings
in~$\Sigma^*$ where $\Sigma$~is a finite alphabet set, is a set of
production rules of the form $V\mapsto W_1W_2$ or $V\mapsto a$, where
$V,W_1,W_2$ are elements of a set~$\mathcal{V}$ of `variables', i.e.,
nonterminal symbols, and~$a\in\Sigma$.  There is a distinguished start
symbol $S\in\mathcal{V}$ with which the process begins.  Applying the
production rules repeatedly yields a subset $L\subset\Sigma^*$, i.e., a
language.  An SCFG assigns probabilities (which add to unity) to the
productions of each $V\in\mathcal{V}$, and yields a probability
distribution over the strings in $L\subset\Sigma^*$, i.e., over~$\Sigma^*$.

The probability distribution $P:\Sigma^*\to[0,1]\subset\mathbb{R}_+$
produced by an SCFG is an example of an $\mathbb{R}_+$-algebraic series.
In general, if $A$~is a semiring, an $A$-algebraic series (of CNF type)
over an alphabet~$\Sigma$ is a weighting function $f\colon\Sigma^*\to A$
obtained as one component (i.e., the component~$f_S$) of the formal
solution of a coupled set of quadratic equations
\begin{displaymath}
  f_V=\sum_{W_1,W_2\in\mathcal{V}} c_{V;W_1,W_2}\, f_{W_1}f_{W_2} +
  \sum_{a\in\Sigma} c_{V;a}\,a,\qquad V\in\mathcal{V},
\end{displaymath}
computed by iteration~\cite{KuichSalomaa}.  The coefficients
$c_{V;W_1,W_2}$ and~$c_{V;a}$ are elements of~$A$, so each~$f_V$ is a sum
of $A$-weighted strings in~$\Sigma^*$, or equivalently a function
$f_V\colon\Sigma^*\to A$.  It~is clear that Theorem~\ref{thm:normalization}
has an analogue: any probability distribution on~$\Sigma^*$ produced by a
SCFG of CNF type is simply a {\em normalized\/} $\mathbb{R}_+$-algebraic
series.

Any SCFG of CNF type has $\left|\mathcal{V}\right|^3 +
\left|\mathcal{V}\right|\left|\Sigma\right|$ parameters, which may be too
many for practical estimation if a small sequence family is being modeled.
To~facilitate modeling, one should use an SCFG with a restricted structure,
and also exploit results from weighted automata theory.  If the nucleotide
distribution does not vary much along typical sequences, then the alphabet
set~$\Sigma$ can be taken to be a $2$-letter alphabet $\{a,b\}$ (if~one is
modeling Watson--Crick pairing, exclusively) or even a $1$-letter alphabet
(if~one is modeling runs of unpaired bases).  Also, one can leverage the
fact that $A$-algebraic series subsume $A$-rational series, which implies
(in~the $1$-letter case) that $A$-algebraic {\em sequences\/}, which are
effectively indexed by~$\mathbb{N}$, subsume $A$-rational sequences.  In
the Boolean ($A=\mathbb{B}$) case, the first statement is the familiar
Chomsky hierarchy.

In the case of a $1$-letter alphabet $\Sigma=\{a\}$, an SCFG defines a
probability distribution on $\mathbb{N}\cong\{a\}^*$.  That is, it defines
an $\mathbb{N}$-valued random variable~$\tau$, the length of the string
emitted by the stochastic push-down automaton (SPDA) corresponding to
the~SCFG.  The SPDA uses~$\mathcal{V}$, the set of nonterminal symbols, as
its stack alphabet, and its stack is initially occupied by the start
symbol~$S$.  The stochastic production rules specify what happens when a
symbol $V\in\mathcal{V}$ is popped off the stack: either two symbols
$W_1,W_2\in\mathcal{V}$ are pushed back, or a letter~`$a$' is emitted.  By
construction, at~least one letter must be emitted by a CNF-type SCFG before
its stack empties, so $Pr(\tau=0)=0$.

The class of probability distributions on~$\mathbb{N}$ associated to SCFGs
(whether or~not of CNF type), i.e., that of normalized
$\mathbb{R}_+$-algebraic sequences, is potentially useful in parametric
stochastic modeling, but has not been widely employed.  It will be denoted
$\mathcal{F}_{\rm alg}$ here, since each distribution in~it has an
algebraic $z$-transform $G(z)=\sum_{n=0}^\infty z^n\,Pr(\tau=n)$.  For any
SCFG, an algebraic equation satisfied by~$G(z)$ can be computed by
polynomial elimination (e.g., by computing the resultant of the above
system of quadratic equations).  Let ${\it PH}_d$ denote the class of
discrete phase-type distributions.

\smallskip
\begin{theorem}
\label{thm:hadamardetc}
  (i) ${\it PH}_d\subset\mathcal{F}_{\rm alg}$.  (ii)~If $X,Y$ are
  independent $\mathbb{N}$-valued random variables~(RVs) with distributions
  in~${\it PH}_d$, then conditioning on~$X=Y$ yields an~RV with
  distribution in~$\mathcal{F}_{\rm alg}$.  (iii)~If, furthermore, $Z$~is
  an independent $\mathbb{N}$-valued RV with distribution
  in~$\mathcal{F}_{\rm alg}$, then conditioning on~$X=Z$ yields an~RV with
  distribution in~$\mathcal{F}_{\rm alg}$.
\end{theorem}

\smallskip
These are `normalized' versions of standard facts on $A$-rational
and~$A$-algebraic series, in particular on their composition under the
Hadamard product $(x_n),(y_n)\mapsto (x_ny_n)$, in the special case when
$A=\mathbb{R}_+$ and~$\Sigma=\{a\}$.  (See~\cite{Fliess74,KuichSalomaa}.)
They have direct probabilistic proofs.  E.g., to prove~(i), one would show
that starting from the distribution of~$\tau\in\mathbb{N}$, the absorption
time in a~HMM, one can construct an SCFG that yields the same distribution
on~$\mathbb{N}$.  (If $Pr(\tau=0)>0$ then the SCFG cannot be of CNF type.)
The procedure is similar to constructing a PDA that accepts a specified
regular language.

Much as with discrete PH distributions, it is difficult to parametrize
distributions in the class~$\mathcal{F}_{\rm alg}$ without, rather
explicitly, parametrizing the stochastic automata (SCFGs or SPDAs) that
give rise to them; or at~least their $z$-transforms.  It is difficult,
in~general, to characterize when a probability distribution on~$\mathbb{N}$
that has an algebraic $z$-transform lies in~$\mathcal{F}_{\rm alg}$.

The following example illustrates the problem.  Any distribution $n\mapsto
Pr(\tau=n)$ on~$\mathbb{N}$ that has an algebraic $z$-transform necessarily
satisfies a finite-depth recurrence of the form $\sum_{k=0}^N
C_k(n)\,Pr(\tau=n+k)=0$, where the functions~$C_k$, $k=0,\dots,N,$ are
polynomial in~$n$.  (If~none of the~$C_k$ depends on~$n$, then the
$z$-transform will be rational.)  Consider, for example, the $2$-term
recurrence
\begin{displaymath}
(n+a)(n+b)\,Pr(\tau=n) = (n+c)(n+1)\,Pr(\tau=n+1),
\end{displaymath}
where $a,b,c\in\mathbb{R}$ are parameters, which is of this form.  The
$z$-transform $G(z)=\sum_{n=0}^\infty z^n\,Pr(\tau=n)$ of its solution is
proportional, by definition, to ${}_2F_1(a,b;c;z)$, which is Gauss's
parametrized hypergeometric function.  The set of triples
$(a,b;c)\in\mathbb{R}^3$ that yields an {\em algebraic\/} $z$-transform,
and hence an $\mathbb{R}$-algebraic sequence $n\mapsto Pr(\tau=n)$, is
explicitly known.  It was derived in the nineteenth century by
H.~A. Schwartz~\cite[Chap.~VII]{Poole36}.  Unfortunately, it is an {\em
infinite discrete\/} subset of~$\mathbb{R}^3$, not a continuous subset.

In~general, the $z$-transform of the solution of a finite-depth recurrence
of the above form will be algebraic in~$z$ only if the overall parameter
vector of its coefficients, the polynomials $\{C_k(n)\}_{k=0}^N$, is
confined to a submanifold of positive codimension.  For distributions
in~$\mathcal{F}_{\rm alg}$, this makes recurrence-based parametrization
less useful than SCFG-based or $z$-transform-based parametrization.

\section{Modeling Secondary Structure}
\label{sec:modeling}
A new scheme for modeling the prior distribution of secondary structures in
an RNA sequence family will now be proposed.  It will exploit the insights
of Sections \ref{sec:duration} and~\ref{sec:algebraic}, on the class of
discrete phase-type distributions on~$\mathbb{N}$ (i.e.,~${\it PH}_d$), and
the larger class of $\mathbb{R}_+$-algebraic distributions on~$\mathbb{N}$
(i.e.,~$\mathcal{F}_{\rm alg}$).

If $\Sigma=\{A,U,G,C\}$ is the alphabet set, any SCFG, or its associated
SPDA, will define a probability distribution on~$\Sigma^*$, the set of
finite length sequences~\cite{Sakakibara94}.  (The distribution of the
sequence length, which is a random variable, lies in~$\mathcal{F}_{\rm
alg}$.)  But even if the SCFG is in Chomsky normal form (CNF), the number
of its parameters grows cubically in the number of grammar variables, as
mentioned above.  To~facilitate estimation, the model should have a
restricted structure.

The models of Knudsen and Hein~\cite{Knudsen99} and Nebel~\cite{Nebel2004}
are representative.  The Knudsen--Hein SCFG has variable set
$\mathcal{V}=\{S,L,F\}$, and production rules
\begin{align*}
  S&\mapsto LS\mid L,{\rm\ i.e.,\ }S\mapsto L^+\defeq L\mid L^2\mid L^3\mid\cdots,\\
  L&\mapsto s\mid a_1Fb_1,\\
  F&\mapsto a_2Fb_2\mid LS,{\rm\ i.e.,\ }F\mapsto a_2Fb_2\mid LL^+.
\end{align*}
Here $s$ signifies an unpaired base and $a_i\dots b_i$ signifies two bases
that are paired in the secondary structure, so $L^+$~produces runs of
unpaired bases, i.e., loops (which may include stems) and $F$~produces runs
of paired bases, i.e., stems (which may include loops of length at
least~$2$).  This SCFG is not a CNF one, but model parameters may be
estimated by a variant of the Inside--Outside algorithm.  If one takes
single base frequencies and pair frequencies (i.e., the probability of
$a_i\dots b_i$ representing $A$\textendash\nobreak$U$,
$G$\textendash\nobreak$C$, or even $G$\textendash\nobreak$U$) into account,
one has only three independent parameters to be estimated, one probability
per production rule.  Knudsen and Hein (cf.\ Nebel) used as their primary
training set a subset of the European database of long subunit ribosomal
RNAs (LSU rRNAs)~\cite{DeRijk98,Wuyts2001}.  For the probabilities of $LS$
vs.~$L$ (from~$S$), they estimated $87\%$ vs.~$13\%$; for $s$ vs.~$a_1Fb_1$
(from~$L$), $90\%$ vs.~$10\%$; and for $a_2Fb_2$ vs.~$LS$ (from~$F$),
$79\%$ vs.~$21\%$.  Their training set actually included tRNAs as~well,
since they were attempting to model the family of folded RNA molecules as a
whole.

As Knudsen and Hein note, their model yields loops and stems with
geometrically distributed lengths.  To improve quantitative agreement, it
would need to be made more sophisticated.  It would also benefit from a
cleaner separation between its two levels: the paired-base and
unpaired-base levels, i.e., the context-free and regular levels (in~the
formal language sense), i.e., the SPDA and HMM levels (in~the stochastic
automata-theoretic sense).  The above production rules couple the two
levels together.  It is not clear from Ref.~\cite{Knudsen99} how well the
model stochastically fits the length of (i)~training sequences, (ii)~the
subsequences comprising paired bases, and (iii)~the subsequences comprising
unpaired bases.  Separating the two levels should facilitate the separate
fitting of these quantities.

By definition, folded RNA secondary structure is characterized by a
subsequence comprising paired bases, so the stochastic modeling of
secondary structure in a given family should initially focus on such
subsequences.  If pseudo-knots (a~thorny problem for automata-theoretic
modeling) are ignored, these subsequences are effectively {\em Dyck
words\/}, or balanced parenthesis expressions.  In the absence of
covariation, one expects to be able to generate such words over
$\{A,U,G,C\}$ from classical Dyck words over the $2$-letter alphabet
$\{a,b\}$, consisting of opening and closing parentheses, by replacing each
$a$\textendash\nobreak$b$ pair independently by an
$A$\textendash\nobreak$U$, $G$\textendash\nobreak$C$, or
$G$\textendash\nobreak$U$ pair, according to observed pair frequencies.
Knudsen and Hein note that order matters: in
$G$\textendash\nobreak$C$\,/\,$C$\textendash\nobreak$G$ pairs in~tRNA,
the~$G$ tends to be nearer the $5'$~end of the RNA than the~$C$.  Still to
be resolved, of~course, is the selection of the underlying probability
distribution over Dyck words in~$\{a,b\}^*$.

One could start with any CFG that unambiguously generates the Dyck words
over $\{a,b\}$, and make it stochastic by weighting its productions.  The
simplest such CFG is a $1$-variable one, $S\mapsto ab\mid abS\mid aSb \mid aSbS$.
The corresponding SCFG is
\begin{displaymath}
S   \mapsto p_1\cdot ab+p_2\cdot abS+p_3\cdot aSb +p_4\cdot aSbS,
\end{displaymath}
where $\sum_ip_i=1$.  This SCFG, with $3$~free parameters, is so simple
that it can be studied analytically.  The length of a Dyck word is an
$\mathbb{N}$-valued random variable, the distribution of which lies
in~$\mathcal{F}_{\rm alg}$, with a parameter-dependent, algebraic
$z$-transform.  As was explained in Section~\ref{sec:algebraic}, it is best
to parametrize distributions in~$\mathcal{F}_{\rm alg}$ by the SCFGs that
give rise to them, rather than by explicit formulas or even by the
recurrence relations that they satisfy; and this is an example.

This Dyck model could be made arbitrarily more versatile, since arbitrarily
complicated CFGs that generate the Dyck language over $\{a,b\}$ can readily
be constructed.  One could, for instance, iterate $S\mapsto ab\mid abS\mid
aSb \mid aSbS$ once, obtaining a production rule for~$S$ with
$25$~alternatives on its right-hand side.  Weighting them with
probabilities would yield an SCFG with $24$~independent parameters, which
would be capable of much more accurate fitting of data on an empirical
family of RNA sequences.  In~general, one could choose model parameters to
fit not~only the observed distribution of Dyck word lengths (i.e.,
per-family paired-base subsequence lengths), but also the distribution of
lengths of stems, i.e., runs of contiguous paired bases, which may be far
from geometric.

The preceding discussion of Dyck words formed from paired bases ignored
loops, i.e., runs of unpaired bases.  They are best handled on a second
level of the SCFG.  The simplest production rule for {\em full\/} sequences
would have not $S$, $abS$, $aSb$, $aSbS$ on its right-hand side, but rather
$IaIbI$, $IaIbS$, $IaSbI$, $IaSbS$, where each of the eight~$I$s expands to
a run of unpaired bases.  In the absence of covariation, modeling each run
is a matter of duration modeling.  Starting from an $\mathbb{N}$-valued
random run length, or equivalently a distribution over finite $1$-letter
words, one would generate a run of unpaired bases by replacing each letter
independently by $A,U,G,C$, according to family-specific single base
frequencies.

Each run length is naturally taken to have a distribution in~${\it PH}_d$,
since that will allow the resulting run of bases to be generated by an HMM
(with absorption).  Each run length will be the absorption time in a
finite-state Markov chain, the parameters of which, i.e., transition
probabilities, can be estimated from empirical data.  Geometric
distributions, and generalizations, are appropriate.  It follows from
Theorem~\ref{thm:3} that employing a large Markov chain with a fully
connected transition graph, and hence a number of parameters that grows
quadratically in the number of states, would {\em not\/} be appropriate.
Without loss of generality, each transition graph can be assumed to have no
`cycles within cycles within cycles'.

In this extended (two-level) stochastic model of the secondary structure of
a family of RNA sequences, the sequence length distribution still lies
in~$\mathcal{F}_{\rm alg}$.  That is because a (Dyck-type) SPDA wrapped
around one or more HMMs is still an SPDA, with an SCFG representation.
This observation is similar to the proof of Theorem~\ref{thm:hadamardetc}:
what has been constructed here is simply an SCFG with a special structure,
not given explicitly in Chomsky normal form.  The full set of model
parameters could be estimated by the Inside--Outside
algorithm~\cite{Lari90}, rather than by estimating Dyck-SCFG and run-length
parameters separately; but that would not be so efficient.

The test of the proposed SCFG architecture will be its value in secondary
structure prediction, since from any RNA sequence the most likely parse
tree, and paired-base subsequence, can be computed by maximum a~posteriori
estimation.



\end{document}